# A Minimal Experimental Bias on the Hydrogen Bond Greatly Improves *Ab Initio* Molecular Dynamics Simulations of Water

Paul B. Calio,[1] Glen M. Hocky,[1,2] and Gregory A. Voth[1,]*

[1]Department of Chemistry, Chicago Center for Theoretical Chemisty, James Franck Institute, and Institute for Biophysical Dynamics, The University of Chicago, 5735 South Ellis Avenue, Chicago, Illinois 60637, United States

[2]Department of Chemistry, New York University, New York, NY, 10003, United States

## Abstract

Experiment Directed Simulations (EDS) is a method within a class of techniques seeking to improve molecular simulations by minimally biasing the system Hamiltonian to reproduce certain experimental observables. In a previous application of EDS to *ab initio* molecular dynamics (AIMD) simulation based on electronic density functional theory (DFT), the AIMD simulations of water were biased to reproduce its experimentally derived solvation structure. In particular, by solely biasing the O-O pair correlation functions, other structural and dynamical properties that were not biased were improved. In this work, the hypothesis is tested that directly biasing the O-H pair correlation (and hence the O-H…H hydrogen bonding), will provide an even better improvement of DFT-based water properties in AIMD simulations. The logic behind this hypothesis is that for most electronic DFT descriptions of water the hydrogen bonding is known to be deficient due to anomalous charge transfer and over polarization in the DFT. Using recent advances to the EDS learning algorithm, we thus train a minimal bias on AIMD water that reproduces the O-H radial distribution function derived from the highly accurate MB-pol model of water. It is then confirmed that biasing the O-H pair correlation alone can lead to improved AIMD water properties, with structural and dynamical properties in even closer to experiment than the previous EDS-AIMD model.

# Introduction

*Ab initio* molecular dynamics (AIMD) simulation[1] has become a popular tool to understand water (see, e.g., refs[2-5]) and aqueous solutions/environments (see, e.g., refs[6-7]). By using AIMD, as opposed to classical empirical water models, electronic structure calculations – primarily within the Kohn-Sham density functional theory (DFT) framework – can be used to solve for the electronic ground state. Electron densities can then respond to the surrounding electric field and thus account for polarization of the electron cloud, and forces can be calculated on-the-fly to propagate the dynamics of the nuclei. Furthermore, AIMD becomes particularly valuable for simulating systems while accounting for chemical reactivity, as it does not require specifying a defined bonding topology, and as such allows one to study dynamics of aqueous systems containing excess protons or hydroxide ions (see, e.g., refs[8-13]).

Accurately modeling the hydrogen bond between water molecules has proven challenging for AIMD water simulations that use generalized gradient approximations (GGA) such as PBE[14] or BLYP[15-16] as the exchange-correlation functional in the DFT.[17] GGA functionals result in over-polarization due to having a small energy gap between Kohn-Sham virtual orbitals and occupied orbitals,[18-19] and they also exhibit partial covalency (or charge-transfer) within the intermolecular interactions of water molecules.[17, 20-21] These inaccuracies in modeling the hydrogen bond result in a water model that is overstructured at room temperature and diffusion coefficients that can be an order or orders of magnitude slower than found in experiment.[9, 22-23]

It is common practice in the AIMD community to mitigate the deficiencies of GGA functionals by increasing the simulation temperature or by going beyond GGA functionals.[2-3, 5, 24] It has previously been shown for the PBE functional that simulation temperatures of 400 K are necessary to mimic room temperature structure and dynamics of water.[23, 25-26] It has been the case that increasing the temperature is not only an *ad hoc* remedy for the glassy behavior of GGA functionals, but additionally used to mimic nuclear quantum effects in GGA AIMD in an *ad hoc* fashion.[27] By contrast, going beyond simple GGA functions and using hybrid or meta-GGA functionals has been shown to improve the hydrogen bond properties and hence the water properties,[2, 24] but doing so significantly increases the computational cost. It should also be noted that it is still commonplace to combine non-GGA functionals and increased simulation temperatures in AIMD as an *ad hoc* fix for nuclear quantum effects in room temperature water.[2-3]

One way to improve the accuracy of MD simulations (including AIMD) is to add a biasing potential to an observable of the system in order to improve agreement between that observable and one in target system. Pitera and Chodera showed using the method of Lagrange multipliers that there exists a linear bias on observables that minimizes the relative entropy of an ideal probability distribution for an observable to a target experimental one.[28] Using this idea, White and Voth developed the Experiment Directed Simulation (EDS) method as a means to parameterize these unknown linear terms in the Hamiltonian on-the-fly within a single MD simulation using a stochastic gradient descent algorithm.[29] Since then, there have been additional advances to EDS and similar methods,[30-31] and these have been recently reviewed by Amirkulova and White.[32] In related work, the Coarse-Grained Directed Simulation (CGDS) method was developed where coarse-grained observables in a molecular sub-system are biased based on data from coarse-grained simulations of a larger supramolecular complex.[33] In this work, the stochastic gradient descent algorithm of the original EDS method was unable to find coupling constants fast enough due to longer timescale and limited configurations of the CG observables that are otherwise available in isotropic systems such as water. In the CGDS paper, however, several variants of the learning algorithm were developed, where the stochastic gradient descent algorithm was replaced with a gradient descent using the covariance of all deviations of collective variables from target observables, and by a Levenberg-Marquadt minimization. In the studies presented here, we will take advantage of these more efficient methods which have better convergence properties.

In previous work,[34] we developed the idea that EDS could be applied to adjust the solvation structure of a relatively low cost DFT AIMD approach, rather than employing the previously mentioned alternatives of increasing the simulation temperature or going beyond GGA functionals. In that work[34] the solvation structure of BLYP and dispersion corrected BLYP water was biased to reproduce the experimental O-O radial distribution function (RDF). This EDS bias, henceforth known as "EDS-AIMD(OO)", was able to improve the targeted solvation structure of water without needing an increase of simulation temperature or computational cost. Importantly, other properties that were not biased also improved as a consequence of having a better O-O structure. For example, when adding an excess proton to the system, the EDS-AIMD(OO) approach improved the ratio of the hydrated excess proton to water diffusion coefficient and did not disrupt the other properties of the system. To confirm that employing EDS to a more accurate model would not adversely affect our results, the method was also tested on a dispersion corrected functional. It was found that, as expected, the linear bias learned from the gradient descent algorithm for the BLYP-D3 system was smaller than the EDS bias of normal BLYP due to the already better agreement with experiment for BYTP-D3, and all results for the biased EDS-BLYP-D3 model were equivalent or better.

The present work builds on this previous study, but this time by biasing the hydrogen bond (intermolecular O-H coordination properties) as opposed to the oxygen coordination structure of water. We chose to pursue this approach due to the intuitive feeling that the having a correct description of hydrogen bonding in DFT water is the underlying deficiency leading to other poor properties, due to the over-polarization and anomalous charge transfer of GGA functionals. Correcting the hydrogen bond structure with EDS is more difficult than the O-O coordination, due to the anisotropic nature of hydrogen bonding, and the need to distinguish an O-H pair as either covalently bonded or hydrogen bonded. By biasing the hydrogen bond, however, we can more directly target the charge transfer and over-polarization found in GGA water simulations which, among other things, gives rise to its glassy behavior and slow diffusion. In this work, we thus bias BLYP and BLYP-D3 simulations to reproduce the O-H RDF of the highly and demonstrably accurate MB-pol water model,[35-38] which includes many-body interactions parameterized from high level CCSD(T) electronic structure calculations and many-body polarization effects. MB-pol accurately reproduces many properties of water and is gaining recognition as one of the most – if not the most – accurate water models available. By using the classical MB-pol O-H RDF as the refence in EDS-AIMD, we are therefore able to effectively include higher order correlations into the BLYP-level AIMD water simulations.

The remainder of this paper is organized as follows: In the Methods section, we first include a subsection that reviews the theory regarding the EDS method and how we apply it to bias the hydrogen bond. In the next subsection of Methods, we describe the details of the EDS algorithm as used, while in the Simulations Procedure subsection, we describe the simulation setup and details for the EDS learning algorithm and the production runs. In the Results and Discussion section, we then present the results and give further discussion. We lastly present conclusions and the outlook for the use of experimental-bias methods in AIMD.

## Methods
### Experiment Directed Simulation Method

EDS modifies the system Hamiltonian by adding a biasing potential that is parameterized to reproduce a target observable.[29] In the present case, the hydrogen bond is biased to reproduce the O-H radial distribution function (RDF) of the MB-pol water model at 298 K. As in the previous work, the RDF is corrected by biasing its statistical moments such as the coordination number (the zeroth moment). The bias on each moment is a function of three variables: a coupling constant, $\alpha_k$, a function characterizing the pairwise distances between all hydrogen-oxygen pairs, $f_k(\vec{r}_{ij})$, and its target average value, $\hat{f}_k$. The EDS potential for each hydrogen atom is therefore:

$$V(r_i) = \sum_{k=0}^{M} \frac{\alpha_k}{\hat{f}_k} f_k(\vec{r}_{ij}) = \sum_{k=0}^{M} \frac{\alpha_k}{\hat{f}_k} \sum_{j=1}^{n_O} r_{ij}^k [1 - u(r_{ij} - r_0)], \tag{1}$$

where M represents the number of moments being biased, $n_o$ is the number of oxygen atoms in the system, and we take $f_k(\vec{r}_{ij})$ to be a product of the O-H pairwise distance and mollified step function. The step function for biasing the hydrogen bond was chosen as:

$$1 - u(r_{ij} - r_0) = \begin{cases} \dfrac{1 - \left(\dfrac{r_{ij} - r_0}{w}\right)^6}{1 - \left(\dfrac{r_{ij} - r_0}{w}\right)^{12}}, & r_{ij} > r_0 \\ 1, & r_0 \geq r_{ij} > r_b \\ 0, & r_{ij} \leq r_b \end{cases} \tag{2}$$

where $r_0$ was set to 2.125 Å, $w$ was set to 0.7, and $r_b$ was set to 1.2 Å. This choice of step function was motivated by the desire to bias the hydrogen bond O-H interactions and not the covalent bond O-H interactions. We set the EDS bias to be zero for all hydrogen-oxygen pairwise distance less than $r_b$ = 1.2 Å, which is where the O-H RDF is zero. The $r_0$ was set to 2.125 Å to bias the trough between the first and second intermolecular peaks of the O-H RDFs peak as this gave final RDFs that best agreed with MB-pol. MB-pol's target values are its coordination numbers and moments determined by integrating MB-pols O-H RDF (Eq. 3).

$$\hat{f}_k = \rho \int_0^\infty dr_{ij} \left[1 - u(r_{ij} - r_0)\right] 4\pi r_{ij}^{2+k} g_{OH}(r_{ij}) \tag{3}$$

In Eq. 3, ρ is the number density, $1 - u(r_{ij} - r_0)$ is the same mollified unit-step function as Eq. 2, and $g_{OH}(r_{ij})$ is the target O-H radial distribution function. *Note that by $k^{th}$ moment we are specifically referring to the power in Eq. 3, and the zeroth moment corresponds to the coordination number.*

Determining EDS Coupling Constants

The determination of EDS coupling constants to bias the hydrogen bond characterized by these AIMD potentials was nontrivial and required improvements to the EDS learning algorithm to improve the rate of convergence of EDS parameters and collective variables (CVs).[33] EDS learns the coupling constants during the progression of a simulation where the coupling constants at time segment $\tau + 1$ are updated similar to a stochastic gradient descent algorithm

$$\alpha_{\tau+1} = \alpha_\tau - \eta_\tau \left(\Delta \alpha_\tau \cdot \left(\frac{\partial \Delta \alpha_\tau}{\partial \alpha_\tau}\right)\right) = \alpha_\tau - \eta_\tau \delta_\tau, \tag{4}$$

where $\boldsymbol{\eta}_\tau$ is the learning rate, $\boldsymbol{\delta}_\tau$ is the step size, and $\Delta_k \alpha_\tau = \langle f_k(r)_\tau \rangle - \widehat{f}_k$, and bold font represent vectors. Each time segment is set for a predefined period of simulation timesteps by the user, and should be about double the autocorrelation time of its corresponding collective variable.

Our current implementation of EDS uses the learning rate of White and Voth, and is calculated as

$$\eta_\tau^k = \frac{A_k}{\sqrt{\sum_{j=1}^{\tau}(\delta_j^k)^2}}, \qquad (5)$$

where $A_k$ is the maximum value a coupling constant can change. The simplest step size uses a derivative term that is proportional to the covariance of the collective variable error at time $\tau$. With $f_i = \langle f_i(r)_\tau \rangle - \widehat{f}_i$:

$$\left(\frac{\partial \Delta \boldsymbol{\alpha}_\tau}{\partial \boldsymbol{\alpha}_\tau}\right)_{ij} = -\text{Cov}_\tau(f_i, f_j) = -\langle f_i f_j \rangle + \langle f_i \rangle \langle f_j \rangle_{ij} \equiv \bar{\bar{J}}_{ij} \qquad (6)$$

In the other case, the Levenberg-Marquardt (LM) algorithm was used for the step-size, where the derivative term in Eq. 4 is replaced with:

$$\left(\frac{\partial \Delta \boldsymbol{\alpha}_\tau}{\partial \boldsymbol{\alpha}_\tau}\right) = [\bar{\bar{J}}^\text{T} \bar{\bar{J}} + \lambda_\tau \text{diag}(\bar{\bar{J}}^\text{T} \bar{\bar{J}})]^{-1} \bar{\bar{J}}^\text{T} \qquad (7)$$

where $\text{diag}(\bar{\bar{J}}^\text{T} \bar{\bar{J}})$ is a purely diagonal matrix with elements of $\bar{\bar{J}}^\text{T} \bar{\bar{J}}$, and $\lambda_\tau$ is known as a mixing parameter which can tune the style of the step size. In this work, the LM algorithm is used to determine the coupling constants for EDS-BLYP and EDS-BLYP-D3 AIMD simulations.

Table 1: Algorithm Parameters and Simulations Settings

| System | Alg. | Init Bias (Coordination, 2nd Moment, kcal/mol) | Target Values (Coordination 2nd Moment, Å$^2$) | A ($k_BT$) | Period (fs) | EDS Sim Length (ps) |
|---|---|---|---|---|---|---|
| **BLYP 1** | LM : $\lambda = 0.1$ | (0,0) | (1.62, 3.86) | 300 | 25 | 50 |
| **BLYP 2** | LM : $\lambda = 0.1$ | (0,0 | (1.62, 3.86) | 300 | 25 | 90 |
| **BLYP 3** | LM : $\lambda = 0.1$ | (0,0) | (1.62, 3.86) | 100 | 25 | 40 |
| **BLYP-D3 1** | LM : $\lambda = 0.1$ | (0,0) | (1.62, 3.86) | 200 | 25 | 100 |
| **BLYP-D3 2** | LM : $\lambda = 0.1$ | (0,0 | (1.62, 3.86) | 200 | 25 | 120 |
| **BLYP-D3 3** | LM : $\lambda = 0.1$ | (0,0) | (1.62, 3.86) | 100 | 25 | 25 |

## Simulation Procedure

EDS was used to bias the coordination and 2nd moment of the O-H (hydrogen bond) RDF. EDS was applied to three independent trajectories of 128 water molecules in a cubic simulation box with box length set to 15.64 Å. All AIMD simulations were done with the quick-step module in CP2K[39] version 3.0 with a modified version of PLUMED2[40] based on version 2.5 that we are putting up on the Voth Group GitHub account using Goedecker-Teter-Hutter (GTH) pseudopotentials[41], a TZV2P basis set with a plane-wave cutoff of 400 Rydbergs, and with BLYP[15-16] exchange correlation functional or BLYP with D3 Grimme dispersion corrections,[42-43] with a timestep of 0.5 fs. Coupling constants were determined in the constant NVT ensemble using three Nosé-Hoover chains at 298 K and with a time constant of 3000 cm$^{-1}$. EDS simulations were run until the CVs were determined to be converged, see Table 1 for details. After this point, the coupling constants were fixed and simulations continued in the constant NVT simulations for 10-15 ps to equilibrate the system with the fixed coupling constant. Finally, production EDS-AIMD runs in the constant NVE ensemble were performed for 80 ps and 40 ps for BLYP and BLYP-D3 simulations, respectively. As a comparison, single BLYP and BLYP-D3 AIMD simulations were run in the constant NVE ensemble for 80 ps and 40 ps, respectively. All dynamical and static properties were obtained from simulations in the constant NVE ensemble and averaged over all three trajectories. Except for the RDFs and unless otherwise noted, our analysis scripts were used to analyze a single MB-pol trajectory consisting of 256 water molecules in the constant NVE ensemble for 50 ps to compare with our results.

## Results and Discussion

By examining Figure 1a, we can see that biasing the hydrogen bond with EDS can greatly improve the O-H RDF of BLYP. In fact, we see the first peak has nearly perfect agreement with the MB-pol reference, and the structural properties past the first peak come into closer agreement with MB-pol. Additionally, by biasing the hydrogen bond in BLYP water, we see greater improvements in the O-O RDF (Fig. 1b) and the H-H RDF (Fig. 1c). As shown in the O-H RDF, we see that the first peak in these other RDFs come into nearly perfect agreement with MB-pol, even though these other properties were not biased but they are still improved due to their strong correlations with the hydrogen bond. We note that the hydrogen bond EDS bias is targeting the close-range interactions between O-H pairs within a hydrogen bond. It should also be noted, however, that EDS greatly improves the structural properties past the first peak in all three RDFs (Fig. 1a-c), although not perfectly.

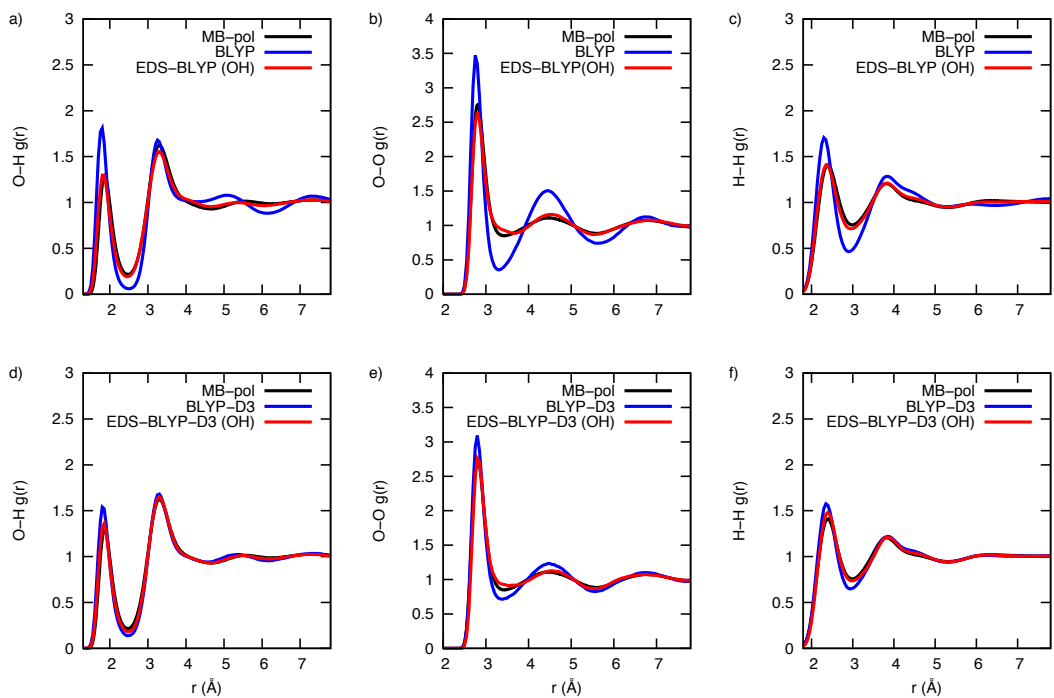

**Figure 1**. Radial distribution function for both BLYP (a-c) and BLYP-D3 simulations (d-f). In black we show the RDFs of the classical MB-pol water model, in blue is either the BLYP or BLYP-D3 RDFs for O-H (a and d), O-O (b and e), or H-H (c and f), and in red is the RDF that results from using EDS to bias the hydrogen bond in BLYP or BLYP-D3 simulations. These RDF are averages over three independent EDS trajectories.

An improvement is also seen in the structural properties of dispersion corrected BLYP (BLYP-D3) over uncorrected BLYP when we bias the hydrogen bond with EDS. Similar to biasing the hydrogen bond in BLYP water simulations, we see that biasing the hydrogen bond in BLYP-D3 water simulations leads to RDFs whose first peaks nearly perfectly reproduces the first peaks in MB-pol (Fig 1d-f). Again, we emphasize that only the hydrogen bond is biased in these EDS-AIMD simulations, and any improvements in other structural properties arise due to strong correlations with the hydrogen bond collective variable. The advantage of biasing BLYP-D3 over BLYP water simulations can be found in the second peaks of the O-H RDF and O-O RDF; where these second peaks reproduce MB-pol's structure. This indicates that EDS is not a substitute for dispersion corrected DFT but complements it.

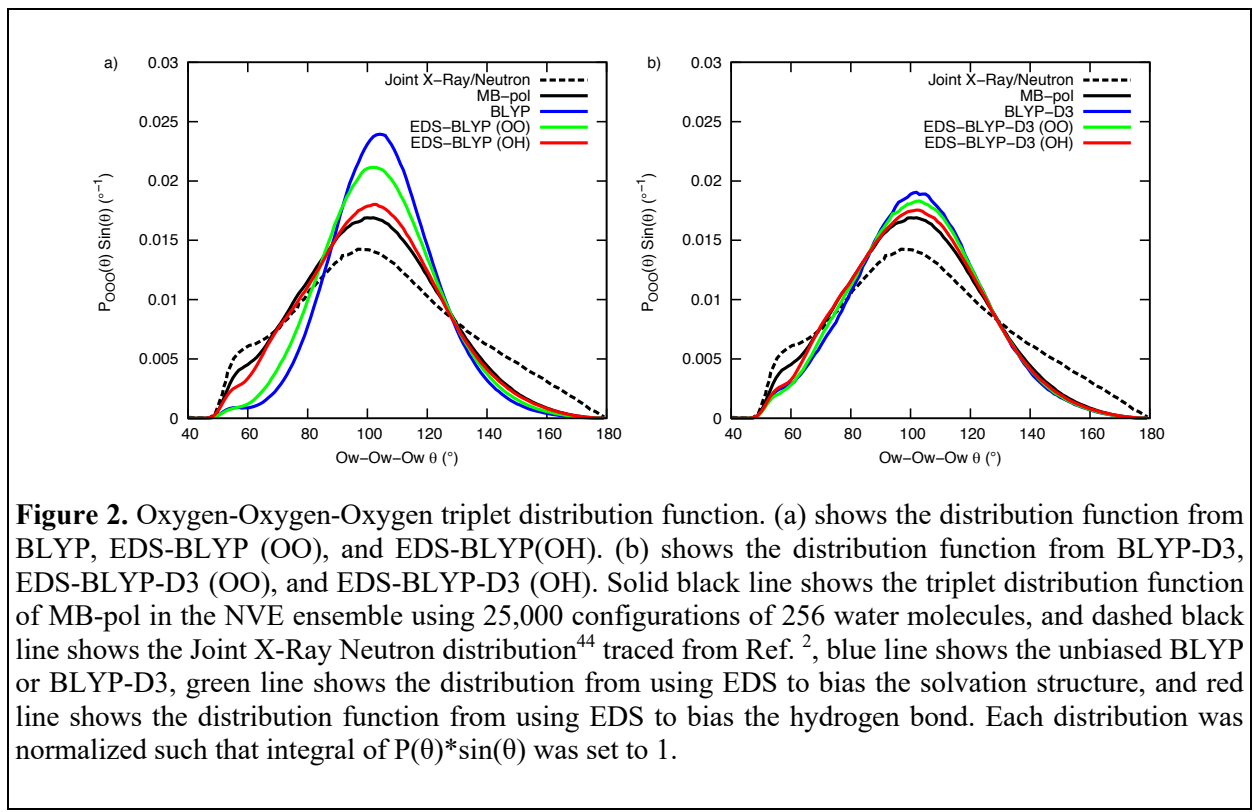

**Figure 2.** Oxygen-Oxygen-Oxygen triplet distribution function. (a) shows the distribution function from BLYP, EDS-BLYP (OO), and EDS-BLYP(OH). (b) shows the distribution function from BLYP-D3, EDS-BLYP-D3 (OO), and EDS-BLYP-D3 (OH). Solid black line shows the triplet distribution function of MB-pol in the NVE ensemble using 25,000 configurations of 256 water molecules, and dashed black line shows the Joint X-Ray Neutron distribution[44] traced from Ref. [2], blue line shows the unbiased BLYP or BLYP-D3, green line shows the distribution from using EDS to bias the solvation structure, and red line shows the distribution function from using EDS to bias the hydrogen bond. Each distribution was normalized such that integral of $P(\theta)*\sin(\theta)$ was set to 1.

We further examined the structural properties by calculating the oxygen-oxygen-oxygen (O-O-O) triplet probability function to characterize the tetrahedrality of the water structure. For a central oxygen atom, all 3-body oxygen angles were calculated for O-O pairwise distances less than 3.22 Å, which is the radial distance where the coordination number is ~ 4. We see that in both the BLYP (Fig. 2a) and BLYP-D3 (Fig. 2b) cases, the unbiased simulations have a large probability above 100°, which is typically attributed to a strong tetrahedral ordering. By using EDS to bias the simulations, we see a decrease in both the biased BLYP and BLYP-D3 plots that is very close to MB-pol, with the most prominent improvement arising from biasing the hydrogen bond (OH) instead of the solvation structure (OO). Also, the peak around 55° in Figure 2a is captured with the OH bias [EDS-BLYP(OH)] but the solvation structure bias (EDS-BLYP(OO)) is unable to capture it. However, MB-pol does an overall better job at reproducing the experimental peak around 55° as MB-pol is explicitly parameterized to reproduce 3-body interactions. Nevertheless, by biasing the hydrogen bond in BLYP and BLYP-D3 simulations, as opposed to biasing the solvation structure (the O-O RDF), we can obtain O-O-O triplet distribution that are in better agreement with MB-pol and experimental data, despite the fact that the O-O-O distribution was not biased directly. We note that nuclear quantum effects are not included in any of these EDS-AIMD or MB-pol results shown here, but they are of course present in the experimental data.

**Table 2**: Final Results for EDS-AIMD Simulations. For comparison, we show dynamical values for MB-pol, EDS-BLYP(OO), BLYP, and BLYP-D3 trajectories. MB-pol diffusion coefficients were taken from Ref. [38]. EDS-BLYP(OO) diffusion coefficients were taken from Ref. [34]. $\vartheta_D$ is the standard deviation for diffusion constant. BLYP time constants show a mono-exponential decay even when fitting with a bi-exponential. $\tau_1$ and $\tau_2$ are time constants obtained via bi-exponential fits of the hydrogen bond ACD (Fig. 3).

| Simulation | $\alpha_0/f_0$ (kcal/mol) | $\alpha_2/f_2$ (kcal/mol Å²) | D (Å²/ps) | $\vartheta_D$ (Å²/ps) | $\tau_1$ (ps) | $\tau_2$ (ps) | Ave. Temp (K) |
|---|---|---|---|---|---|---|---|
| MB-pol | - | - | 0.23 | 0.02 | 0.41 | 4.95 | - |
| BLYP | - | - | 0.01 | 0.01 | - | 20.22 | 317.32 |
| EDS-BLYP(OO) | - | - | 0.06 | 0.02 | 0.85 | 11.61 | - |
| EDS-BLYP(OH) 1 | 387.80 | -70.53 | 0.139 | 0.002 | 0.31 | 6.11 | 290.48 |
| EDS-BLYP(OH) 2 | 421.52 | -69.88 | 0.167 | 0.001 | 0.58 | 6.06 | 297.872 |
| EDS-BLYP(OH) 3 | 299.57 | -60.94 | 0.186 | 0.003 | 0.33 | 4.75 | 302.96 |
| BLYP-D3 | - | - | 0.06 | 0.05 | 0.79 | 13.45 | 302.88 |
| EDS-BLYP-D3(OH) 1 | 285.87 | -36.93 | 0.152 | 0.008 | 0.12 | 5.38 | 298.82 |
| EDS-BLYP-D3(OH) 2 | 237.26 | -35.16 | 0.118 | 0.003 | 0.59 | 7.49 | 294.80 |
| EDS-BLYP-D3(OH) 3 | 218.75 | -31.59 | 0.146 | 0.002 | 0.32 | 6.40 | 293.06 |

Another goal in biasing the hydrogen bond in BLYP and BLYP-D3 AIMD water simulations was to see an improvement in the dynamics. The self-diffusion coefficient of BLYP water has been reported to be an order (or orders) of magnitude slower[9, 22-23] than that found in experiment (0.23 Å$^2$/ps).[45] In the AIMD simulations reported here, the BLYP water oxygen self-diffusion coefficient was found to be 0.01 $\pm$ 0.01 Å$^2$/ps, and with an improvement to 0.06 $\pm$ 0.05 Å$^2$/ps for the dispersion corrected BLYP-D3 water simulations. The unbiased BLYP-D3 water diffusion is the same as using EDS to bias the O-O solvation structure of BLYP water [EDS-BLYP(OO)] at 0.06 $\pm$ 0.02 Å$^2$/ps.[34] (The diffusion coefficient of EDS-BLYP-D3(OO) water was not calculated in that work.) On the other hand, biasing the hydrogen bond (O-H RDF) in EDS-BLYP(OH) and EDS-BLYP-D3(OH) water improves the self-diffusion coefficient to be 0.164 $\pm$ 0.004 Å$^2$/ps, and 0.139 $\pm$ 0.009 Å$^2$/ps, respectively. This is much closer to experiment (0.23 Å$^2$/ps), and significantly larger than using the O-O RDF solvation structure for EDS bias.[34] (Note that the addition of nuclear quantum effects would likely increase the EDS-AIMD diffusion results even more, bringing them into even better agreement with experiment.) One can also confirm that the increased diffusion is not a result of a spurious increase in simulation temperature (see Table 2).

The increased diffusion coefficient of EDS-BLYP(OH) and EDS-BLYP-D3(OH) over BLYP, dispersion corrected BLYP-D3, and even EDS-BLYP(OO) results from directly targeting the hydrogen bond with a bias in BLYP water. This physical improvement is further emphasized by the hydrogen bond autocorrelation function results in Figure 3. The autocorrelation function is taken over all pairs and all times, and a hydrogen bond is defined as having an O-O distance of less than 3.5 Å and an H-O-O angle less than 30°. Biasing the hydrogen bond clearly does a better job at breaking hydrogen bond correlations than does simply biasing the O-O solvation structure;

this behavior additionally contributes to the observed larger diffusion for the O-H EDS biased AIMD simulations over the O-O EDS biased ones (see Table 2).

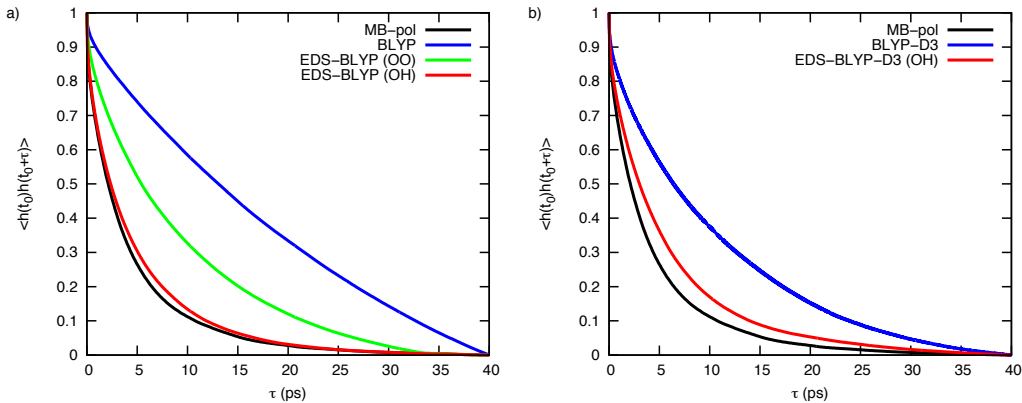

**Figure 3.** Hydrogen Bond autocorrelation function (ACF). (a) show the ACF from MB-pol, BLYP, EDS-BLYP (OO), and EDS-BLYP(OH). (b) shows the distribution function from MB-pol, BLYP-D3, and EDS-BLYP-D3(OH). Black line shows MB-pol, blue line shows the unbiased BLYP or BLYP-D3, green line shows the distribution from using EDS to bias the solvation structure, and red line shows the distribution function from using EDS to bias the hydrogen bond. The hydrogen bond ACF comes from averaging over the first 40 ps of the simulation. The hydrogen bond ACF for EDS-BLYP-D3(OO) was not included due to the more limited statistical sampling in that prior work. Time constants from a bi-exponential fit of these curves can be found in Table 2.

It is valuable and interesting to look at the form of the bias potential learned by the EDS algorithm in order to reproduce the first peak in the O-H RDF (in this case, the MB-pol target result). In our previous implementation of EDS to bias the O-O solvation structure in BLYP and BLYP-D3 water,[34] the EDS biasing potential added an effective repulsive force to the oxygen-oxygen pair interaction, with the EDS-BLYP-D3(OO) adding a smaller force magnitude than the EDS-BLYP(OO) since the former is closer to the target RDF result. In the present cases, since dispersion corrected BLYP-D3 water simulations more closely match the accurate reference MB-pol results, we expect the EDS potential and force for EDS-BLYP-D3(OH) to be a smaller overall in magnitude than for EDS-BLYP(OH).

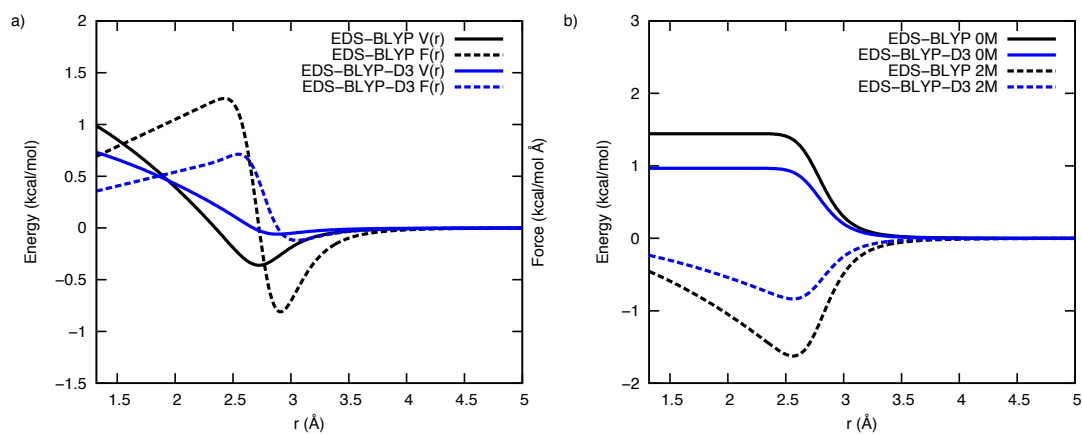

**Figure 4**. Potential energy and force plots for the EDS O-H bias. Black lines are for EDS-BLYP(OH) and blue lines are for EDS-BLYP-D3 OH) simulations. (a) The average potential and force energy for the EDS-BLYP (OH) and EDS-BLYP-D3 (OH). Solid lines indicate the EDS potential while the dashed lines indicate the EDS force. (b) The EDS bias as separated into its $0^{th}$ and $2^{nd}$ moment components. Solid lines are for the $0^{th}$ moment contribution to the EDS potentials, and the dashed lines are for the $2^{nd}$ moment contributions to the EDS potentials.

In Figure 4a we show the EDS potential (solid) and force (dashed) for EDS-BLYP(OH) and EDS-BLYP-D3(OH) simulations. The EDS bias in both cases has a repulsive "ramp" at short distances, which is followed by a potential energy well at larger distances. This indicates that using EDS to bias the hydrogen bond is not as simple as adding a repulsive bias between the pairs as was found when biasing the O-O solvation shell.[34] The EDS-BLYP potential has a sharper repulsive wall and deeper attractive well than the dispersion corrected EDS-BLYP-D3 case, which results in stronger repulsive and attractive force magnitudes. We also separated the EDS potentials into the $0^{th}$ and $2^{nd}$ moment contributions in Figure 4b. It is seen in both moments that the EDS-BLYP potential has larger magnitudes than EDS-BLYP-D3, which adds additional explanation for the differences in the EDS-BLYP and EDS-BLYP-D3 potentials in Fig. 4a.

# Conclusions

In this work, the Experiment Directed Simulation (EDS) method has been implemented to bias the O-H hydrogen bond in BLYP and BLYP-D3 AIMD water to reproduce a target O-H RDF of the highly accurate MB-pol model (taken as the "exact" result). The rationale for this EDS bias is that most GGA-level DFT functionals result in hydrogen bonds that are too strong and hence AIMD water at 298 K that is over structured and much too slowly diffusing. By biasing only the hydrogen bond correlations, we were able to improve other structural properties and provide a very significant improvement in the water self-diffusion coefficient. Biasing the hydrogen bond opposed to the O-O solvation structure directly targets the over-polarization and charge transfer present in GGA DFT to obtain a better AIMD water model, and produces superior dynamical properties as shown in the prior text and Supporting Information. We note that once the EDS bias is determined, this results in the addition of only a classical correction term and hence negligible change to the computational cost of the AIMD.

An overall advantage of AIMD is it can also account for chemical reactions. This becomes particularly useful for simulating the hydrated excess proton or hydroxide, which diffuses through the Grotthuss mechanism.[46-48] In previous work,[34] we also implemented an EDS-AIMD O-O solvation structure bias to simulate AIMD water having an excess proton added to it, making the approximation that such a bias is still appropriate for the water solvent even once an excess proton is introduced into the system. It was found that such an approximation is quite accurate (and it is expected to be even more accurate for any system in which a solute or solutes are chemically more distinct from the water solvent). On the other hand, the same approximation cannot be simply implemented in the O-H EDS bias approach taken in this paper. This is because the EDS-AIMD hydrogen bond bias applies forces to all pairs of hydrogens and oxygens within the specified cutoff distances, but one would not want to apply this bias to any special hydrogens considered as part of a hydronium or hydroxide ion complex. Thus, in future work we plan to rigorously combine the EDS-AIMD hydrogen bond bias into a continuous potential for the hydrated excess proton (by combining EDS with a bond order analysis) in such a way that forces are applied to only select classes of hydrogens and oxygens. Such a result can also likely be generalized to other forms of chemical reactions in EDS-AIMD(OH) water in which the water participates in the chemistry. This research is currently in progress.

**Supporting Information**

Additional Information regarding the choice of biased moments, calculations of the EDS potential and force, and Potential of Mean Force (PMF) and 2D hydrogen bond PMF can be found in the Supporting Information. This material is available free of charge via the Internet at http://pubs.acs.org.

Author Information

Corresponding Author

*E-mail gavoth@uchicago.edu

**Acknowledgements**

This research was supported in part by the U.S. Department of Energy, Office of Basic Energy Sciences, Separation Science Program of the Division of Chemical Sciences, Geosciences, and Biosciences under Award Number DE-SC0018648 and in part by the Office of Naval Research (ONR) through award N00014-18-1-2574. Support for GMH came from an NIH Ruth L. Kirschstein NRSA Fellowship (F32 GM113415) and the New York University College of Arts and Sciences. The authors thank Professor Francesco Paesani for the MB-pol RDFs and MB-pol trajectories. The authors also thank Dr. Thomas Dannenhoffer-Lafage and Mr. Chenghan Li for many insightful discussions. Computations were performed in part using resources provided by the University of Chicago Research Computing Center (RCC) and resources at the Texas Advanced Computing Center provided via the Extreme Science and Engineering Discovery Environment (XSEDE) under grant number TG-MCA94P017.

**For Table of Contents Only**

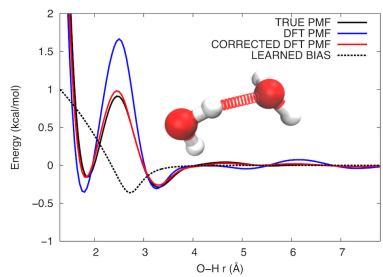

# Supporting Information for:

# A Minimal Experimental Bias on the Hydrogen Bond Greatly Improves *Ab Initio* Molecular Dynamics Simulations of Water

Paul B. Calio,[1] Glen M. Hocky,[1,2] and Gregory A. Voth[1,*]

[1]Department of Chemistry, Chicago Center for Theoretical Chemisty, James Franck Institute, and Institute for Biophysical Dynamics, The University of Chicago, 5735 South Ellis Avenue, Chicago, Illinois 60637, United States

[2]Department of Chemistry, New York University, New York, NY, 10003, United States

## Section S1. Discussion on the Choice of Biased Moments

We used EDS to bias the $0^{th}$ and $2^{nd}$ moment of the O-H RDF in AIMD simulations. In this discussion, we show that it is not necessary to bias the $1^{st}$ moment in these simulations to reproduce the O-H RDF of MB-pol

In Figure S1 we show the $0^{th}$ (biased), $1^{st}$ (unbiased), and $2^{nd}$ (biased) moments in the constant NVE ensemble of the various simulation methods used in this study. We first note that both the $0^{th}$ moment and $2^{nd}$ moment in the EDS simulations are much closer to the target value than the unbiased simulations Additionally, we find the $1^{st}$ moment is much closer to the target value without a direct bias due to the strong correlations between the moments. This seems to suggest that directly biasing the $1^{st}$ moment is not necessary to improve its distribution.

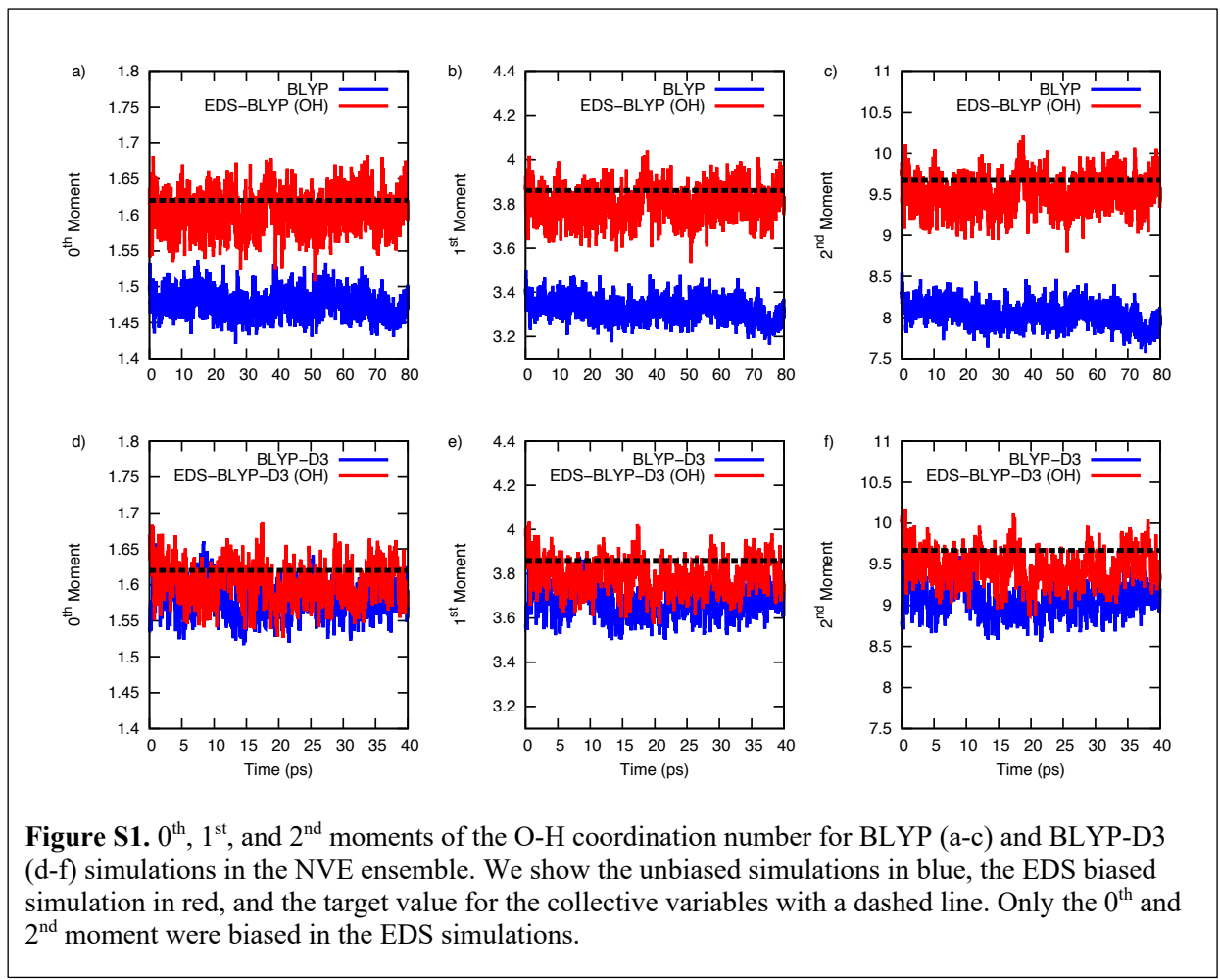

**Figure S1.** $0^{th}$, $1^{st}$, and $2^{nd}$ moments of the O-H coordination number for BLYP (a-c) and BLYP-D3 (d-f) simulations in the NVE ensemble. We show the unbiased simulations in blue, the EDS biased simulation in red, and the target value for the collective variables with a dashed line. Only the $0^{th}$ and $2^{nd}$ moment were biased in the EDS simulations.

We also found that including the $1^{st}$ moment into the EDS learning algorithm had a negligible effect on the resulting EDS potential energy. In Figure S2, we show the EDS-BLYP (OH) potential energy that results from biasing the $0^{th}$ and $2^{nd}$ moment, and the resulting EDS-BLYP (OH) potential after a 30 ps learning simulation where the $0^{th}$, $1^{st}$, and $2^{nd}$ moments were biased. We see excellent agreement between the two potential energy functions except at short distances where there is very little O-H probability. Including the 1st moment only broadens the coupling constant values but does not affect the resulting EDS potential.

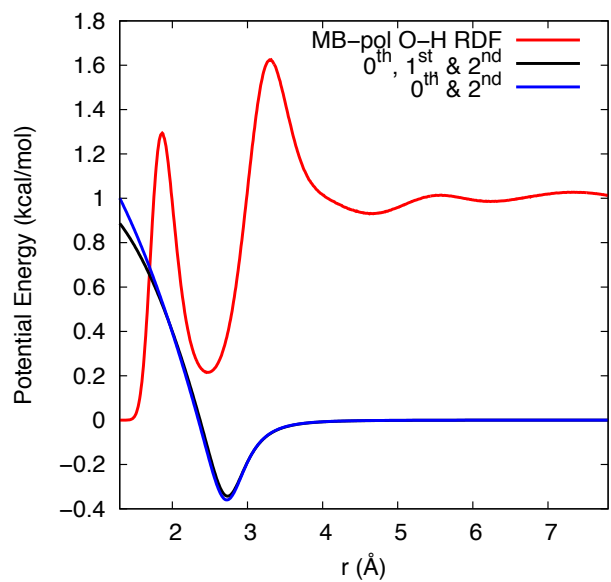

**Figure S2.** EDS potential energy resulting from biasing the $0^{th}$, $1^{st}$, and $2^{nd}$ (black) and the $0^{th}$ and $2^{nd}$ (blue) moments. The MB-pol O-H RDF is present in red.

## Section S2. Calculation of the EDS Potential and Force

The EDS potential for a given pairwise distribution is defined in the main text as the product of the coupling constant ($\alpha_k/f_k$) and the H-O pairwise function. (See Eq. 1). In practice, the PLUMED2 EDS module calculates the mean pairwise distribution over all O-H pairs and subtracts the target value from this mean value. The EDS bias is then the product of this difference and the coupling constant. Although this changes the potential, the forces remain unchanged due to the target values being a constant value. In the main text (Fig 4) and the supporting information (Fig. S4), we chose to present the EDS potential between a single O-H pair using Eq. S1, where we have not subtracted the target value from the pairwise distribution, and have removed any dependence on the number of hydrogen bonds in the system. We show in Fig. S3 the corresponding EDS potential and force plots where the target value is subtracted from the pairwise function. EDS potential plots are the average over all three EDS simulations and the EDS force plots are calculated using finite difference of the potential.

$$V(r_{ij}) = \frac{1}{n_{HBonds}} \sum_{k=0}^{M} \frac{\alpha_k}{\hat{f}_k} \cdot r_{ij}^k [1 - u(r_{ij} - r_0)], \qquad (S1)$$

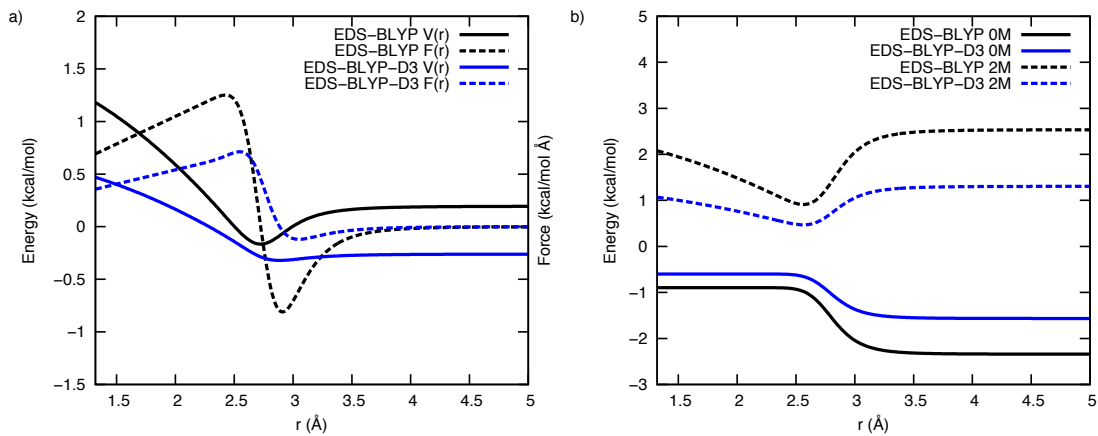

**Figure S3**. Potential energy and force plots for the EDS bias where the target value is subtracted from the pairwise distribution. Black lines are for EDS-BLYP (OH) and blue lines are for EDS-BLYP-D3 (OH). (a) shows that average potential energy and force for the EDS-BLYP (OH) and EDS-BLYP-D3 (OH). Solid lines indicate the EDS potential while the dashed lines indicate the EDS force. (b) The EDS bias is separated into its $0^{th}$ and $2^{nd}$ moment components. Solid lines are for the $0^{th}$ moment of the EDS potential, and dashed lines are for the $2^{nd}$ moment of the EDS potential.

## Section S3. Physical Properties not Described in the Main Test

In this section, we present other important quantities that improved due to the EDS H-bond (O-H) biasing procedure. These are the: potential of mean force (PMF) (S4.1), and hydrogen bond 2D PMF (S4.2). We note that in all cases, biasing the hydrogen bonds using EDS was superior to biasing the oxygens (O-O). In parenthesis of each label signifies where the bias was on the solvation structure (OO) or the hydrogen bond (OH).

**Section S3.1** Potential of Mean Force (PMF) for the O-H and O-O pair-wise distance calculated via equation S2. Here $k_B$ is Boltzmann's constant, $T$ is the temperature of 298 K, and $g(r)$ is the corresponding radial distribution function.

$$F(r) = -k_B T \ln(g(r)) \quad (S2)$$

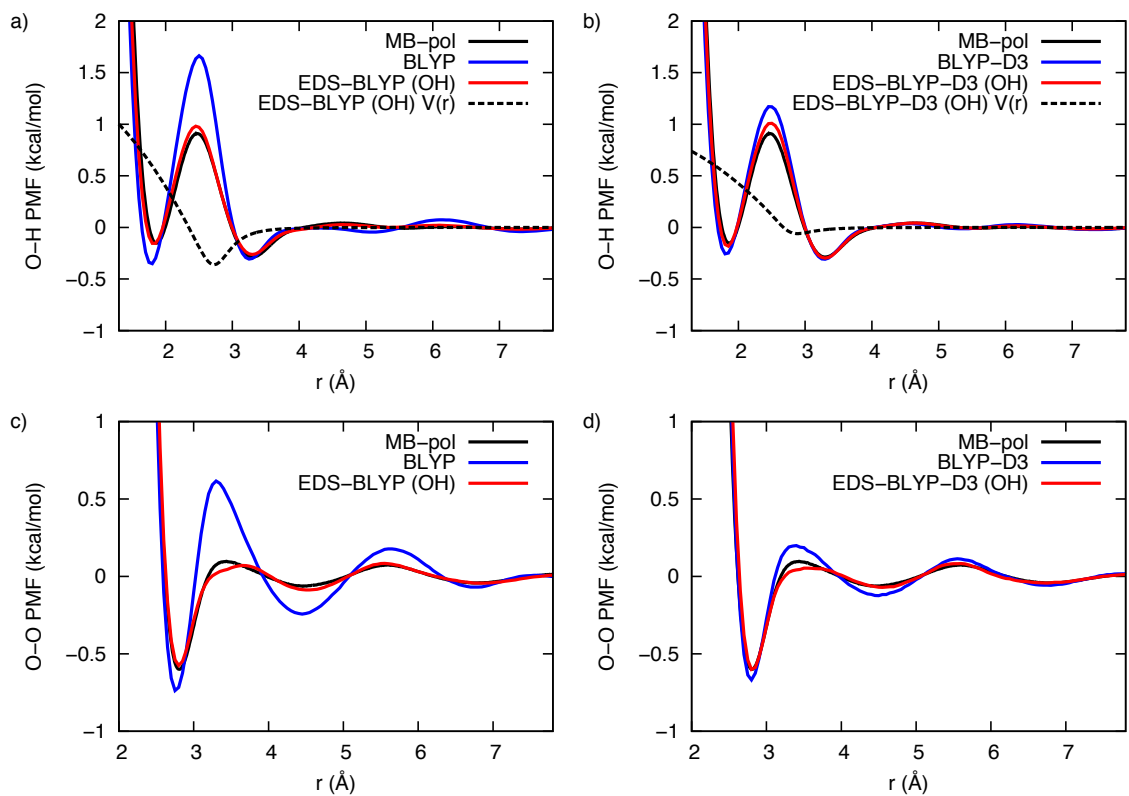

**Figure S4.1** Potential of Mean Force (PMF) calculated from Equation S2. (a) and (b) show the PMF of the O-H pair distribution for BLYP and BLYP-D3 simulations, respectively. (c) and (d) show the PMF of the O-O pair distribution for BLYP and BLYP-D3 simulations, respectively. MB-pol is represented by a solid black line, blue corresponds to BLYP or BLYP-D3 AIMD, red line corresponds to the EDS corrected BLYP or BLYP-D3 AIMD simulations. (a) and (b) additionally show the hydrogen bond EDS bias in dashed black line. The mean EDS bias calculated using the coupling constants in Table I and Equation S1.

**Section S3.2** Hydrogen-Bond 2D PMF. The free-energy is calculated via Eq. S3, where C is a constant chosen to set the minimum of the free-energy to 0 kcal/mol. By using EDS to bias the hydrogen bond, we see a broadening in the hydrogen-bond 2D PMF compared to the unbiased simulations.

$$F(r) = -k_B T \ln(P(\theta,r)) + C \qquad (S3)$$

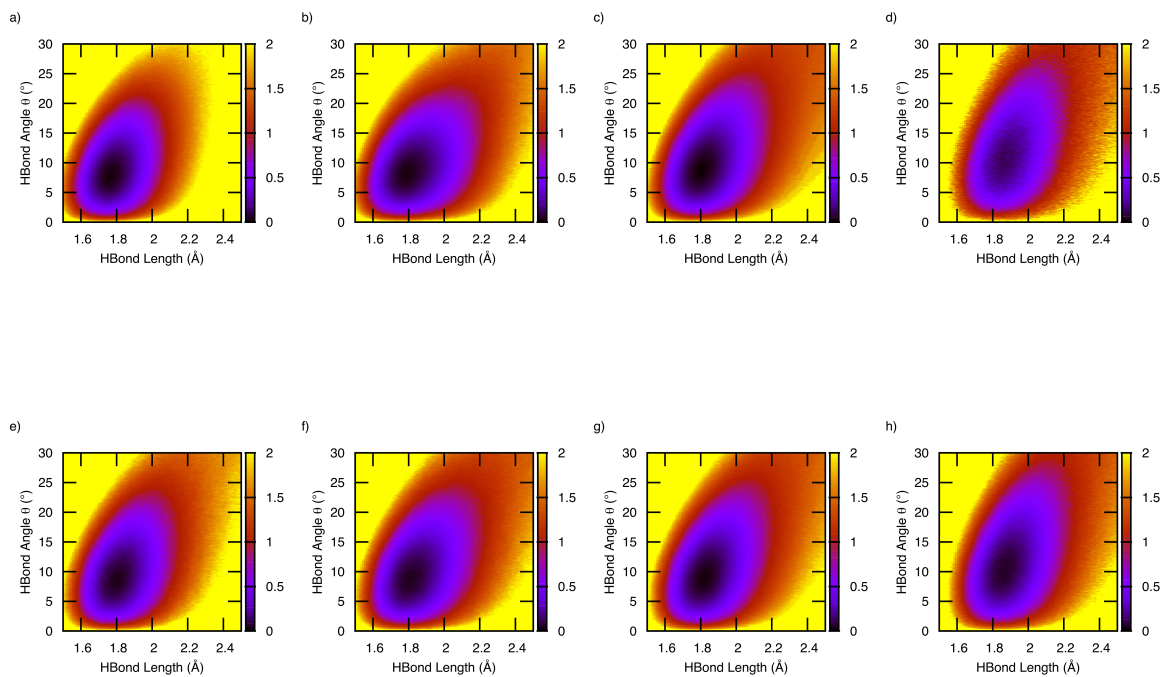

**Figure S4.2** Hydrogen Bond Potential of Mean Force (PMF) using equation S2 for (a) BLYP, (b) EDS-BLYP(OO), (c) EDS-BLYP(OH), (d) MB-pol in NVT ensemble, (e) BLYP-D3, (f) EDS-BLYP-D3 (OO), (g) EDS-BLYP-D3(OH), and (h) MB-pol in the NVE ensemble. The PMF is in units of kcal/mol.